\documentclass[conference]{IEEEtran}
\IEEEoverridecommandlockouts
\usepackage{cite}
\usepackage{amsthm,amsmath,amssymb,amsfonts}
\usepackage{graphicx}
\usepackage{textcomp}
\usepackage{commath}
\usepackage{booktabs}
\usepackage{hyperref}
\usepackage[dvipsnames]{xcolor}         % colors
\definecolor{citecol}{HTML}{6F130C}
\definecolor{tableofcontent}{HTML}{1F4A83}
\definecolor{urlcol}{HTML}{2470D8}

\usepackage{hyperref}
\hypersetup{
    colorlinks=true,       % false: boxed links; true: colored links
    linkcolor=tableofcontent,          % color of internal links (change box color with linkbordercolor)
    citecolor=citecol,        % color of links to bibliography
    urlcolor=urlcol,           % color of external links
}
\usepackage{algorithmic}
\newtheorem{theorem}{Theorem}

\newtheorem{lemma}[theorem]{Lemma}
\newtheorem*{remark}{Remark}

\newcommand{\rd}{\,\mathrm{d}}
\newcommand{\bsx}{\boldsymbol{x}}

\def\BibTeX{{\rm B\kern-.05em{\sc i\kern-.025em b}\kern-.08em
    T\kern-.1667em\lower.7ex\hbox{E}\kern-.125emX}}
\begin{document}

\title{Adaptive Importance Sampling and Quasi-Monte Carlo Methods for 6G URLLC Systems\\
\author{Xiongwen Ke, Houying Zhu, Kai Yi, Gaoning He, Ganghua Yang, Yu Guang Wang}
\IEEEcompsocitemizethanks{\IEEEcompsocthanksitem X. Ke and K. Yi are with The University of New South Wales, Australia; H. Zhu is with Macquarie University, Australia; 
G. He and G. Yang are with Huawei Technologies, Co. Ltd., Shanghai, China; 
Y. G. Wang is with Shanghai Jiao Tong University, China.
Email: yuguang.wang@sjtu.edu.cn} }

\maketitle

\begin{abstract}
 In this paper, we propose an efficient simulation method based on adaptive importance sampling, which can automatically find the optimal proposal within the Gaussian family based on previous samples, to evaluate the probability of bit error rate (BER) or word error rate (WER). These two measures, which involve high-dimensional black-box integration and  rare-event sampling, can characterize the performance of coded modulation. We further integrate the quasi-Monte Carlo method within our framework to improve the convergence speed. The proposed importance sampling algorithm is demonstrated to have much higher efficiency than the standard Monte Carlo method in the AWGN scenario.

\end{abstract}


\section{Introduction}
In wireless communication, reliability is assessed by measuring the correct rate of packets of a certain size delivered within a specified time. For a single user, $N$ packets can be specified to be sent according to business requirements, the packet transmission delay can be measured, and reliability can be evaluated by recording whether the packet transmission delay is within the specified time. 
The reliability (or success rate) of packet transmission in traditional mobile broadband (MBB) systems (2G/3G/4G) is usually at the level of 0.1 or 0.01. This is because they mainly serve for voice and video transmission between humans who are relatively delay tolerant.
In addition to MBB, 5G/6G introduces URLLC systems that are designed for communications between ``intelligence'' including humans and machines which ask for higher reliability at the level of $10^{-6}$ or even to $10^{-9}$ \cite{tong20216g}. This enables many future applications such as telesurgery, robot cooperation, and remote control of machines.
However, the efficiency of evaluating a high-reliability system is becoming a crucial problem since generating error packets at very low probability becomes a rare event and therefore performing Monte-Carlo simulation is time-consuming.
For example, suppose the reliability target value is 99.999\% (32bytes @ $x$ ms), assuming that 100,000 packets are sent, if the delay when transmitting 99,999 packets is less than or equal to $x$ ms, then the reliability of transmitting 32-byte packets within $x$ ms is 99.999\%. For the whole system, the required statistic is the proportion of users who satisfy 99.999\% of the total number of simulated users. Then, $10^6$ sampling points are needed to find the incorrect packets with traditional sampling methods, which would exert an impractical big cost.

Importance sampling (IS) technology provides a remedy, which can reduce the variance of the MC estimator by tiling or scaling the original target distribution as a proposal. The target function in practice is usually complex, like modulated codewords with noise.
Using importance sampling, the resulting biased distribution would have a higher frequency of occurrence of rare events. This biased distribution will be corrected by using the importance weight. The optimal proposal distribution is not known since it depends on the black-box integral function. In fact, even if the optimal proposal distribution is known, it is usually difficult to sample. Thus, the proper choice of the family of proposal distribution and its parameter is crucial to find a sub-optimal proposal. For a gentle introduction to the importance of sampling in communications systems, see \cite{smith1997quick}.

Two families of distributions have often been used as proposals in the literature on communications systems. \cite{romano2013minimum,mahadevan2007snr} used the Bernoulli distribution as a proposal  for importance sampling to evaluate the performance of binary linear block codes. \cite{romano2013minimum} also developed an adaptive approach to update the parameters in the Bernoulli proposal. \cite{font2019importance} used the Gaussian proposal for importance sampling to estimate the random coding Union (RCU) bound and gave the asymptotic analysis of the estimator. However, the RCU bound \cite{polyanskiy2010channel} is only the theoretical maximal channel coding
rate achievable at a given block length and error probability, which is a bit far away from practice.  The Gaussian proposal has also been used to evaluate the performance of Low-density Parity Check (LDPC) Codes in an AWGN channel\cite{cavus2009low} and Rayleigh Fading Channels\cite{ahn2013evaluation}. However, both of them didn't discuss how to determine the parameter for the Gaussian proposal.

\begin{figure}[t]
\centering
\includegraphics[width=\columnwidth]{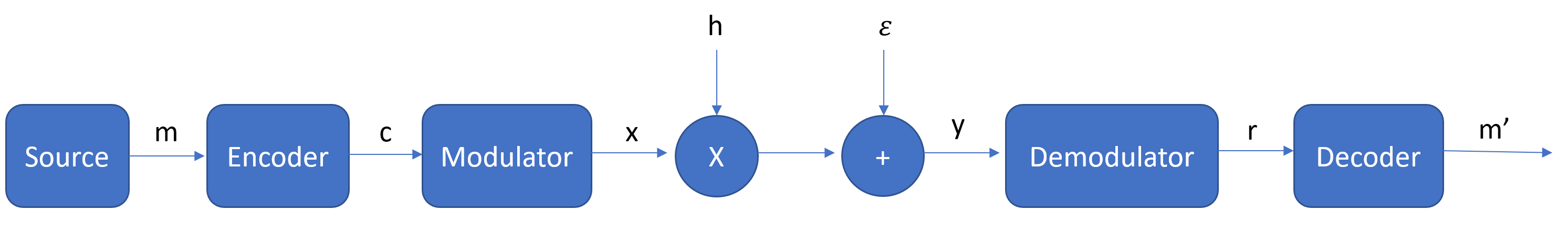}
\caption{Illustration of system model}
\label{fig:system model}
\end{figure}

Our paper makes two main contributions. First, we introduce an adaptive importance sampling scheme based on the work of Bugallo et al. \cite{bugallo2017adaptive} to update the parameters of our Gaussian proposal. Second, instead of relying on traditional random number generators, we use Quasi-Monte Carlo (QMC) and randomized QMC methods to sample from our Gaussian proposal. QMC and RQMC offer a substitute for the widely utilized Monte Carlo method in scientific computing and financial analysis \cite{Niederreiter1992, Leobacher2014, Lemieux2009}. Unlike Monte Carlo, which uses a random sampling approach, QMC uses a deterministic rule to generate points. 
As a result, QMC can achieve the same level of integration error with fewer points, leading to increased efficiency in the integration process, thus higher efficiency in importance sampling.  In this paper, we focus on additive white Gaussian noise (AWGN) in link-level simulation when the encoder and decoder are omitted, see Figure~\ref{fig:system model}. 

% \begin{itemize}
%     \item IS for proposal by adaptive
%     \item work for Gaussian noise (AWGN)
%     \item MC methods
%     \item QMC methods
%     \item for 6G WC
%     \item what results? faster convergence/then more efficient sampling
% \end{itemize}

\section{Model setting and performance measure}

Let $m\in \mathcal{X}^{k}$ be the message vector. A codeword $c\in \mathcal{X}^{n}$ is obtained by encoding this message word. The signal $X$ is the sequence after modulation of codeword $c$. At the output of the channel, a word $r\in \mathcal{X}^{n}$  perturbed by some noise before demodulation, is what we can observe in practice. The decoder will try to recover the observation $r$ to the original message $m$. In communication system modelling, we can have the binary phase-shift keying (BPSK) modulation with additive white Gaussian noise (AWGN) channels or Rayleigh fading channels. For example, in Figure \ref{fig:system model}, when we set $h\equiv1$ and $\epsilon$ as Gaussian distribution, we obtain the AWGN channel.

The performance of the codes is measured by some metric, averaged over all words $r$  generated by some independent drawings from some joint distribution $\pi(Y|X)\pi(X)$, and then transferred by the demodulator. For example, by averaging the 0-1 loss (i.e. if the received word is erroneously decoded, then the loss is 1 and 0 otherwise), we get the probability of word error (WER). In general, we can write 
% \begin{equation}
    $P_{e}=E_{\pi(Y)}[I(Y)] \approx \frac{1}{N}\sum_{i=1}^{N}I(y_{i}),$
% \end{equation}
where $y_{i}\sim \pi(Y)$. Due to the complex structure of both the demodulator and decoder and the length of the code, we cannot explicitly write down the active area of indicator function $I(Y)$. We need to use Monte Carlo simulation to evaluate this high-dimensional integration. When the SNR is high, $P_{e}$ could be very small. For the case of WER, the relative error can be approximated by
% \begin{equation}
    $\frac{\sqrt{Var[\hat{P}_{e}]}}{P_{e}}\approx\frac{1}{\sqrt{P_{e}N}},$
% \end{equation}
which indicates that, for rare event sampling, a large sample size is required to maintain accuracy.
\section{Importance sampling}
To sample the rare event, a good candidate solution is importance sampling (IS)
\begin{equation}
\hat{P}_{e}\approx \frac{1}{N}\sum_{i=1}^{N}I(y_{i})\frac{\pi(y_{i})}{g(y_{i})}=\frac{1}{N}\sum_{i=1}^{N}I(y_{i})w(y_{i}),
\end{equation}
where $w(y_{i})=\frac{\pi(y_{i})}{g(y_{i})}$, $y_{i}\sim g(Y)$ and $g(Y)$ is the density of proposal distribution, which has higher probability to incur the error. There also exists another estimator for IS, which only requires both the proposal and target distribution known up to a normalizing constant by normalizing the importance weight:
% \begin{equation}\label{eq:Pe}
    $\hat{P}_{e}\approx \frac{\sum_{i=1}^{N}I(Y_{i})w(y_{i})}{\sum_{i=1}^{N} w(y_{i})}.$
% \end{equation}
Note that $\pi(Y)=\int \pi(Y|X)\pi(X)dx$ has no closed form in practice. For example, in AWGN channel, $\pi(Y|X)$ is the pdf of $\prod_{i=1}^{n} N(y_{i},\sigma^{2})$ and $X$ can follow the discrete uniform distribution. Then, we have two strategies:
\begin{enumerate}
    \item We jointly sample $Y$ and $X$ based on the proposal $g(Y,X)=g(Y|X)\pi(X)$. In this case the importance weight change to $w(y_{i},x_{i})=\frac{\pi(y_{i}|x_{i})}{g(y_{i}|x_{i})}$.
    \item We apply the conditional importance sampling approach \cite{bucklew2005conditional,rowland2020conditional} by first sampling $x_{i}\sim \pi(X)$, then conditional on $x_{i}$, we use importance sampling to evaluate $E_{\pi(Y)}[I(Y)|X]$. In this case, we have the IS estimator
\end{enumerate}
\vspace{-2mm}
   \begin{equation}\label{CIS}
   \begin{aligned}
    P_{e}&=E_{\pi(X)}[E_{\pi(Y|X)}[I(Y)|X]]\\
    &\approx \frac{1}{S}\sum_{j=1}^{S}\frac{\sum_{i=1}^{N}I(Y_{ij})w(y_{ij},x_{j})}{\sum_{i=1}^{N} w(y_{ij},x_{j})}.
    \end{aligned}
    \end{equation}
This estimator will work well if the error is dominated by the noise $\epsilon$ and the codeword $x$ has a minor impact on the error. For example, when the loss is WER, we could have 
\begin{equation}
E_{\pi(Y|X)}[I(Y)^{2}|X]=E_{\pi(Y|X)}[I(Y)|X] \approx p_{e}.
\end{equation}
Then, by the law of total variance, 
\begin{equation}
\begin{aligned}
\mathbf{Var}_{\pi(Y)}[I(Y)] 
& = E_{\pi(X)}[\mathbf{Var}_{\pi(Y|X)}[I(Y)|X]]\\&+\mathbf{Var}_{\pi(X)}[E_{\pi(Y|X)}[I(Y)|X]]\\
& \approx E_{\pi(X)}[\mathbf{Var}_{\pi(Y|X)}[I(Y)|X]]\\
& \approx \mathbf{Var}_{\pi(Y|X)}[I(Y)|X]=p_{e}-p_{e}^{2}.
\end{aligned}
\end{equation}
In other words, the variance of the IS estimator in equation (\ref{CIS}) is close to the case that we have the analytic form of $\pi(y_{i})$. This implies that we can sample $x_{i}$ such that $S \ll N$. In the extreme case, $S=1$.

Conditional on $X$, the optimal proposal is given by
\begin{equation}
g^{*}(Y|X)=\frac{I(Y)\pi(Y|X)}{E_{\pi(Y|X)}[I(Y)|X]},
\end{equation}
which is impossible to sample. The key problem in importance sampling is designing a sub-optimal proposal. We need to balance the following four points.
\begin{enumerate}
    \item The proposal distribution can efficiently sample rare events compared to the target distribution.
    \item The importance weight can be evaluated quickly.
    \item Sampling from the proposal distribution is straightforward.
    \item The effective sample size should not be too small.
\end{enumerate}

\section{Importance sampling with AWGN channel}

If codeword $x$ is sent over the AWGN channel, then the conditional target distribution is $y|x \sim N(x,\sigma^{2}I_{n})$. To tackle the rare event simulation, we consider two approaches to construct the proposal distribution. 

\subsection{Exponential Tilting}
Exponential Tilting is used in Monte Carlo estimation for rare-event simulation, which will shift the target distribution. Suppose $\pi(Y,X)$ is the pdf of normal distribution  $N(x,\sigma^{2}I_{n})$, the tilted density $g_{\theta}(y)$ is $N(x+\theta,\sigma^{2}I_{n})$, where $\theta=(\theta_{1},...,\theta_{n})$ is the vector of tilting parameters we need to tune. Then, the unnormalize importance weight reads
\begin{equation}
    \frac{\pi(y_{i}|x_{i})}{g_{\theta}(y_{i}|x_{i})}= \exp\left[-\frac{(y_{i}-x_{i})^{T}\theta}{\sigma^{2}}\right]\exp\left[\frac{\|\theta\|^{2}}{2\sigma^{2}}\right].
\end{equation}
The resulting IS estimator is
\begin{equation}\label{eq:IS pe}
    \hat{p}_{e} \approx \frac{\sum_{i=1}^{N}\exp\left[-\frac{(y_{i}-x_{i})^{T}\theta}{\sigma^{2}}\right]I(y_{i})}{\sum_{i=1}^{N}\exp\left[-\frac{(y_{i}-x_{i})^{T}\theta}{\sigma^{2}}\right]},
\end{equation}
or if we use conditional importance sampling,
\begin{equation}\label{eq:IS pe cond}
\hat{p}_{e} \approx \frac{1}{S}\sum_{j=1}^{S}\frac{\sum_{i=1}^{N}\exp\left[-\frac{(y_{ij}-x_{j})^{T}\theta}{\sigma^{2}}\right]I(y_{ij})}{\sum_{i=1}^{N}\exp\left[-\frac{(y_{ij}-x_{j})^{T}\theta}{\sigma^{2}}\right]}.
\end{equation}
The  variance of importance sampling is
\begin{equation}\label{eq: variance_IS}
\begin{aligned}
&\mathbf{Var}_{g(Y|X)\pi(X)}\left[I(Y)\frac{\pi(Y|X)}{g(Y|X)}\right]\\
=&E_{\pi(Y|X)\pi(X)}\left[I(Y)\frac{\pi(Y|X)}{g(Y|X)}\right]-p_{e}^{2}, 
\end{aligned}
\end{equation}
which is not possible to obtain. But our problem is to find the tilting parameter $\theta$ to minimize the variance above. Since only the first term of the equation (\ref{eq: variance_IS}) depends on $\theta$, we can find a surrogate objective function, which is the second-moment estimator for IS. 
\begin{lemma}\label{lem:2nd moment}
The second moment estimator for IS in the exponential tilting case is
\begin{equation}\label{eq:exp tilt 2nd moment}
E_{\theta}[\hat{p}_{e}^{2}]=\frac{\sum_{i=1}^{N}I(y_{i})^{2}\exp\left[\frac{-2(y_{i}-x_{i})^{T}\theta+\|\theta\|_{2}^{2}}{2\sigma^{2}}\right]}{N}
\end{equation}
with its derivative 
\begin{equation}\label{eq:derivative 2nd moment}
\begin{aligned}
\frac{\partial E_{\theta}[\hat{p}_{e}^{2}]}{\partial \theta} = & \frac{1}{N}\sum_{i=1}^{N}I(y_{i})^{2}\exp\left [\frac{-2(y_{i}-x_{i})^{T}\theta+\|\theta\|_{2}^{2}}{2\sigma^{2}}\right]\\
& \times \left[\frac{-2(y_{i}-x_{i})+2\theta}{2\sigma^{2}}\right],
\end{aligned}
\end{equation}
where $(x_{i},y_{i})$ is sampled from the target distribution $\pi(Y,X)$. In addition,  $\frac{\partial^{2}  E_{\theta}[\hat{p}_{e}^{2}]}{\partial \theta^{2}}>0$ for all $\theta_{i}$, which implies that the second moment estimator for IS in \eqref{eq:exp tilt 2nd moment} is a convex function with respect to the variable $\theta$.
\end{lemma}

Therefore, we have the following theorem.
\begin{theorem}\label{thm:optimal tilting}
The optimal tilting parameter $\hat{\theta}$ that minimizes the variance of the IS estimator given by \eqref{eq:IS pe} or \eqref{eq:IS pe cond} is
\begin{equation}\label{eq:optimal tilting}
\hat{\theta} =\frac{\sum_{i=1}^{N}I(y_{i})^{2}(y_{i}-x_{i})\exp[-\frac{(y_{i}-x_{i})^{T}\hat{\theta}}{\sigma^{2}}]}{\sum_{i=1}^{N}I(y_{i})^{2}\exp[-\frac{(y_{i}-x_{i})^{T}\hat{\theta}}{\sigma^{2}}]},
\end{equation}
where $(x_{i},y_{i})$ is sampled from $\pi(Y,X)$. Thus, as $N \rightarrow \infty$, 
\begin{equation}\label{eq:asymp optimal tilting}
\begin{aligned}
 \hat{\theta} \rightarrow_{a.s} & \frac{E_{g(Y,X)}\left[I(Y)^{2}(Y-X)\frac{\pi^{2}(Y,X)}{g^{2}(Y,X)}\right]}{E_{g(Y,X)}\left[I(Y)^{2}\frac{\pi^{2}(Y,X)}{g^{2}(Y,X)}\right]}\\
 & \approx \frac{\sum_{i=1}^{N}I(y_{i})^{2}(y_{i}-x_{i})\exp\left[-\frac{2(y_{i}-x_{i})^{T}\hat{\theta}}{\sigma^{2}}\right]}{\sum_{i=1}^{N}I(y_{i})^{2}\exp\left[-\frac{2(y_{i}-x_{i})^{T}\hat{\theta}}{\sigma^{2}}\right]},
\end{aligned}
\end{equation}
where $g(Y|X)$ could be the pdf of $N(x+\theta,\sigma^{2}I_{n})$ for any $\theta$ and $(x_{i},y_{i})$ is sampled from $g(Y|X)\pi(X)$.
\end{theorem}

\begin{remark} 
Lemma~\ref{lem:2nd moment} and equation \eqref{eq:optimal tilting} in Theorem~\ref{thm:optimal tilting} tell us that it is not possible to obtain a closed-form expression for $\hat{\theta}$, but the numerical result can be obtained by using gradient descent with convergence guaranteed by the convexity of $E_{\theta}[\hat{p}_{e}^{2}]$. However, 
evaluating the gradient from equation \eqref{eq:derivative 2nd moment} needs to sample the rare event from the target distribution. When the SNR is high, the Monte Carlo gradient from \eqref{eq:derivative 2nd moment} might be unreliable. A remarkable consequence of equation \eqref{eq:asymp optimal tilting} is that we can obtain an unbiased IS estimator for $\hat{\theta}$ by using any exponential tilting normal distribution. This provides two merits: 1. Sampling the rare event from the tilting distribution is easier. 2. We can update the estimator $\hat{\theta}$ based on samplers from the previous proposal.
\end{remark}

\subsection{Scaling Variance}
Another way to sample the rare event is by scaling the variance of the target distribution. In this case, the scale density $g_{c}(y)$ is $N(x,c\sigma^{2}I_{n})$, where $c>1$ is the scaling parameter to be tuned. The unnormalize importance weight then reads
\begin{equation}
    \frac{\pi(y_{i}|x_{i})}{g_{\theta}(y_{i}|x_{i})}=c^\frac{n}{2}\exp\left[\left(\frac{1}{c}-1\right)\frac{\|y_{i}-x_{i}\|_{2}^{2}}{2\sigma^{2}}\right].
\end{equation}

Similar to exponential tilting, we can obtain the optimal IS estimator using scaling variance.
\begin{lemma}
The second moment estimator for IS in the scaling variance case is
\begin{equation}
E_{c}[\hat{p}_{e}^{2}]=\frac{c^\frac{n}{2}\sum_{i=1}^{N}I(y_{i})^{2}\exp\left[\left(\frac{1}{c}-1\right)\frac{\|y_{i}-x_{i}\|_{2}^{2}}{2\sigma^{2}}\right]}{N}
\end{equation}
with its derivative given by
\begin{equation}
\begin{aligned}
\frac{\partial E_{c}[\hat{p}_{e}^{2}]}{\partial c} & =\frac{n}{2}c^{\frac{n}{2}-1}\frac{\sum_{i=1}^{N}I(y_{i})^{2}\exp\left[\left(\frac{1}{c}-1\right)\frac{\|y_{i}-x_{i}\|_{2}^{2}}{2\sigma^{2}}\right]}{N}\\
&\hspace{-5mm} -c^{\frac{n}{2}-2} \frac{\sum_{i=1}^{N}I(y_{i})^{2}\exp\left[\left(\frac{1}{c}-1\right)\frac{\|y_{i}-x_{i}\|_{2}^{2}}{2\sigma^{2}} \right]\frac{\|y_{i}-x_{i}\|_{2}^{2}}{2\sigma^{2}}}{N},
\end{aligned}
\end{equation}
where $(x_{i},y_{i})$ is sampled from the target distribution $\pi(Y,X)$. 
\end{lemma}

\begin{theorem}\label{thm:optimal scaling var}
By setting $\frac{\partial E_{c}[\hat{p}_{e}^{2}]}{\partial c}=0$, the stationary point $\hat{c}$ satisfies
\begin{equation}
\hat{c} = \frac{2\sum_{i=1}^{N}I(y_{i})^{2}\exp\left[(\frac{1}{c}-1)\frac{\|y_{i}-x_{i}\|_{2}^{2}}{2\sigma^{2}} \right]\frac{\|y_{i}-x_{i}\|_{2}^{2}}{2\sigma^{2}}}{n\sum_{i=1}^{N}I(y_{i})^{2}\exp\left[(\frac{1}{c}-1)\frac{\|y_{i}-x_{i}\|_{2}^{2}}{2\sigma^{2}}\right]},
\end{equation}
where $(x_{i},y_{i})$ is sampled from $\pi(Y,X)$. Thus, as $N \rightarrow \infty$,
\begin{equation}\label{eq:asymp scaling var}
\begin{aligned}
\hat{c} \rightarrow_{a.s} & \frac{2E_{g(Y,X)}\left[I(Y)^{2}\frac{\|Y-X\|_{2}^{2}}{2\sigma^{2}}\frac{\pi^{2}(Y,X)}{g^{2}(Y,X)}\right]}{nE_{g(Y,X)}\left[I(Y)^{2}\frac{\pi^{2}(Y,X)}{g^{2}(Y,X)}\right]}\\[1mm]
\approx &\frac{2\sum_{i=1}^{N}I(y_{i})^{2}\exp\left[(\frac{1}{c}-1)\frac{\|y_{i}-x_{i}\|_{2}^{2}}{\sigma^{2}} \right]\frac{\|y_{i}-x_{i}\|_{2}^{2}}{2\sigma^{2}}}{n\sum_{i=1}^{N}I(y_{i})^{2}\exp\left[(\frac{1}{c}-1)\frac{\|y_{i}-x_{i}\|_{2}^{2}}{\sigma^{2}}\right]},
\end{aligned}
\end{equation}
where $g(Y|X)$ is the pdf of $N(x,c\sigma^{2}I_{n})$ for any $c$ and $(x_{i},y_{i})$ is sampled from $g(Y|X)\pi(X)$.
\end{theorem}

\begin{remark}
    Theorem~\ref{thm:optimal scaling var} is slightly weaker than Theorem~\ref{thm:optimal tilting} because $\frac{\partial^{2}  E_{c}[\hat{p}_{e}^{2}]}{\partial c^{2}}>0$ can not be guaranteed. Hence, $\hat{c}$ in equation \eqref{eq:asymp scaling var} is only the stationary point. In each update, we let $c^{*}=\max\left\{1+\delta,\hat{c}\right\}$, where $\delta>0$. This guarantees that the proposal distribution always has a larger variance than the target distribution.
\end{remark}

\subsection{Adaptive Importance Sampling}
The adaptive importance sampling \cite{bugallo2015adaptive} is an iterative process to generate the proposal densities to reduce the variance of importance sampling. The procedure consists of three basic steps: generating samples from proposals, calculating the importance weight of each sample (weighting), and updating the parameters in the distribution family of the proposal to obtain new proposals for the next iteration (adapting). The way to make adaption is tailored to the specific problem as mentioned. Many methods have been proposed \cite{cappe2004population,cornuet2012adaptive,martino2015adaptive} for combining the samples from different proposals. Here, we choose a straightforward way, which is to normalize the weighted cross of all samples from all proposals.
Algorithm~1 below implements our adaptive importance sampling strategy.
\begin{figure}[h]
    \centering
    \vspace{-6mm}
\includegraphics[width=\columnwidth]{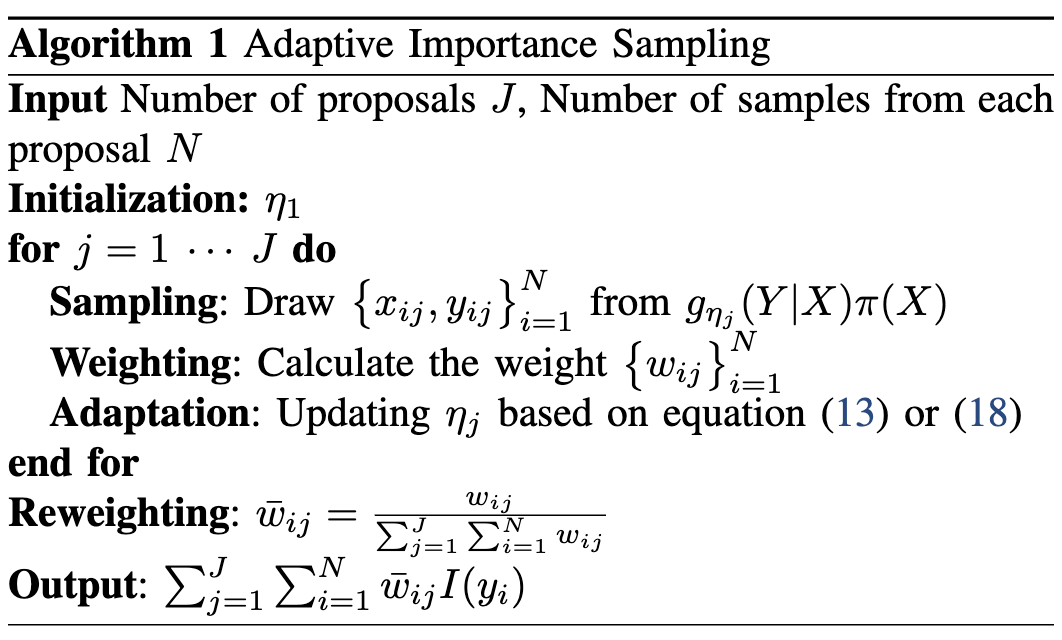}
    % \caption{Caption}
    % \label{fig:my_label}
\end{figure}
% \begin{algorithm}
% \caption{Adaptive Importance Sampling}\label{alg:AIS}
% \State \textbf{Input}  Number of proposals $J$, Number of samples from each proposal $N$
% \State\textbf{Initialization:} $\eta_{1}$
% \For{$j=1 \: \cdots \: J $}
% \State \textbf{Sampling}: Draw $\left\{x_{ij},y_{ij}\right\}_{i=1}^{N}$ from  $g_{\eta_{j}}(Y|X)\pi(X)$ 
% \State \textbf{Weighting}: Calculate the weight $\left\{w_{ij}\right\}_{i=1}^{N}$
% \State \textbf{Adaptation}: Updating $\eta_{j}$ based on equation \eqref{eq:asymp optimal tilting} or \eqref{eq:asymp scaling var}
% \EndFor
% \State \textbf{Reweighting}: $\bar{w}_{ij}=\frac{w_{ij}}{\sum_{j=1}^{J}\sum_{i=1}^{N}w_{ij}}$
% \State \textbf{Output}: $\sum_{j=1}^{J}\sum_{i=1}^{N}\bar{w}_{ij}I(y_{i})$
% \end{algorithm}

\begin{figure*}[t]
    \centering
    \begin{minipage}{0.24\textwidth}
    \includegraphics[width=\textwidth]{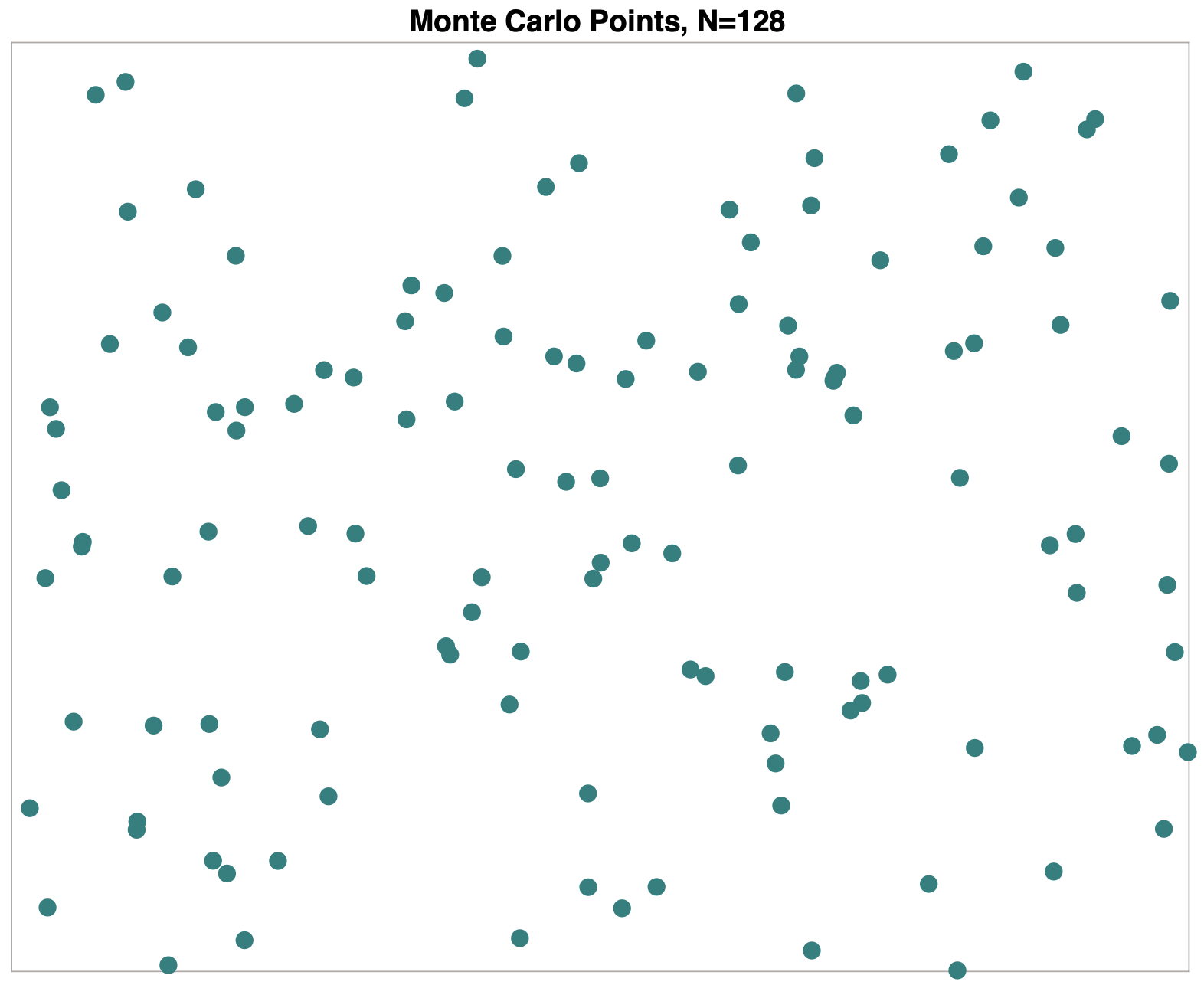}
    \end{minipage}
    \begin{minipage}{0.24\textwidth}
    \includegraphics[width=\textwidth]{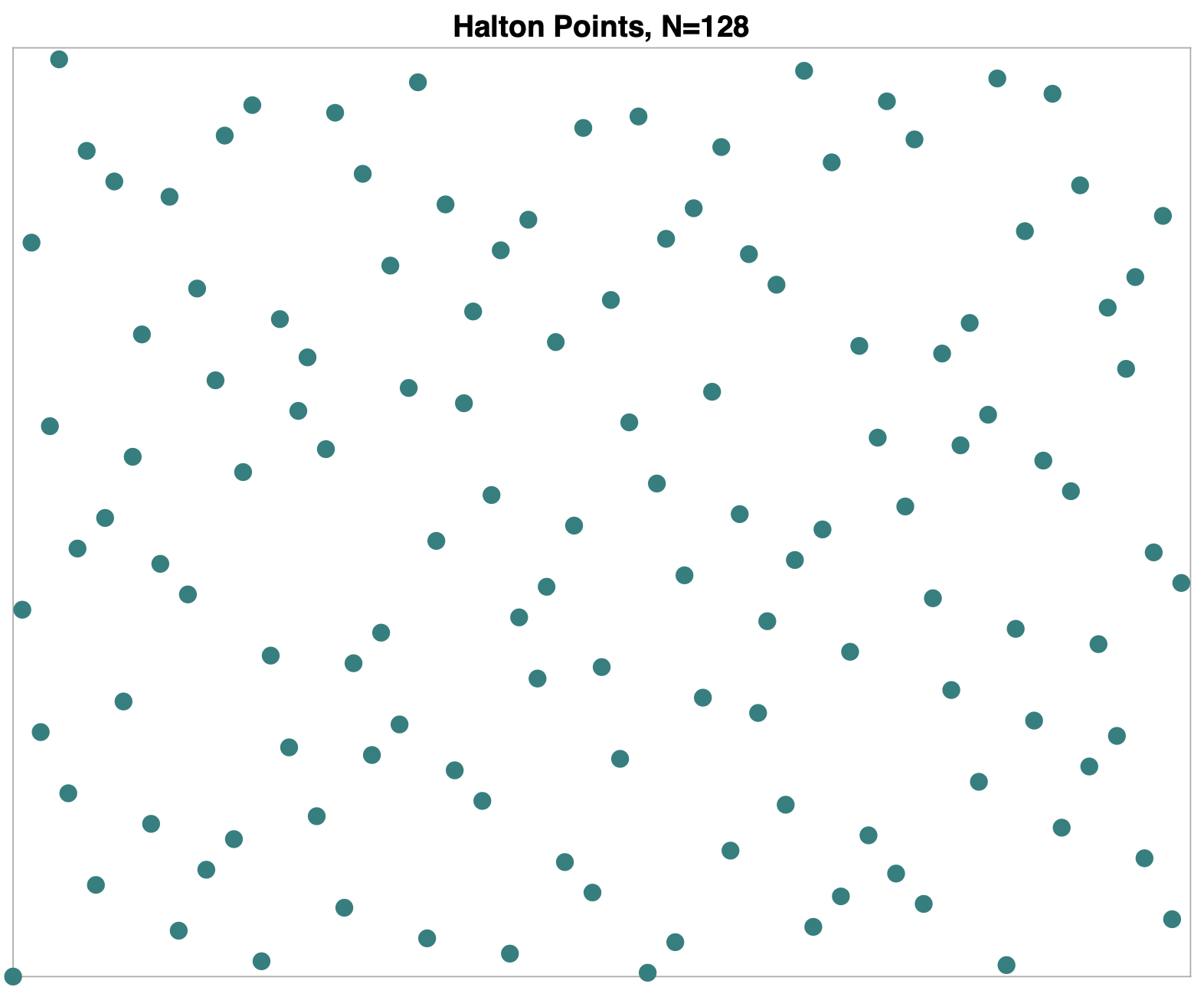}
    \end{minipage}
    \begin{minipage}{0.24\textwidth}
    \includegraphics[width=\textwidth]{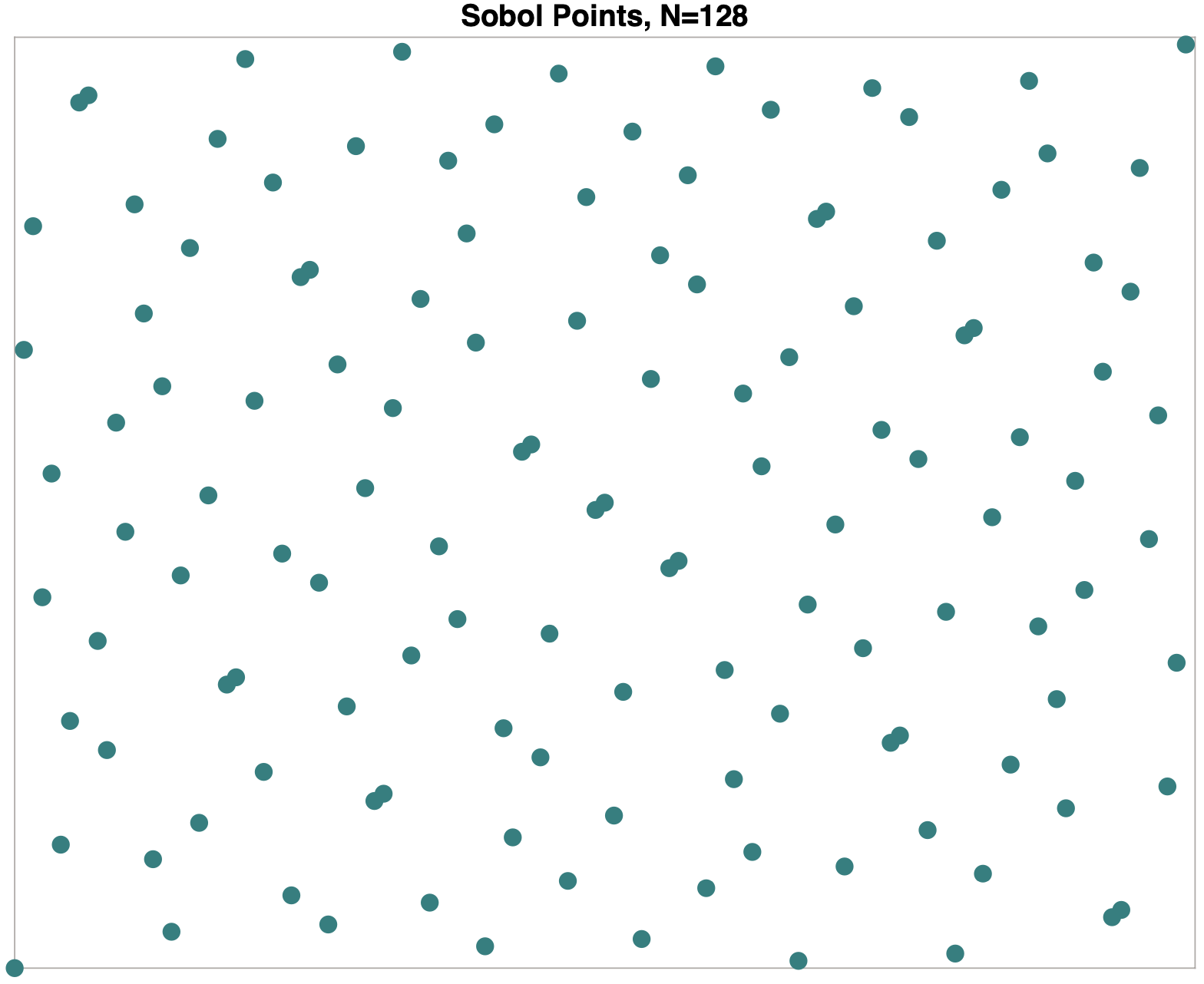}
    \end{minipage}
    \begin{minipage}{0.24\textwidth}
    \includegraphics[width=\textwidth]{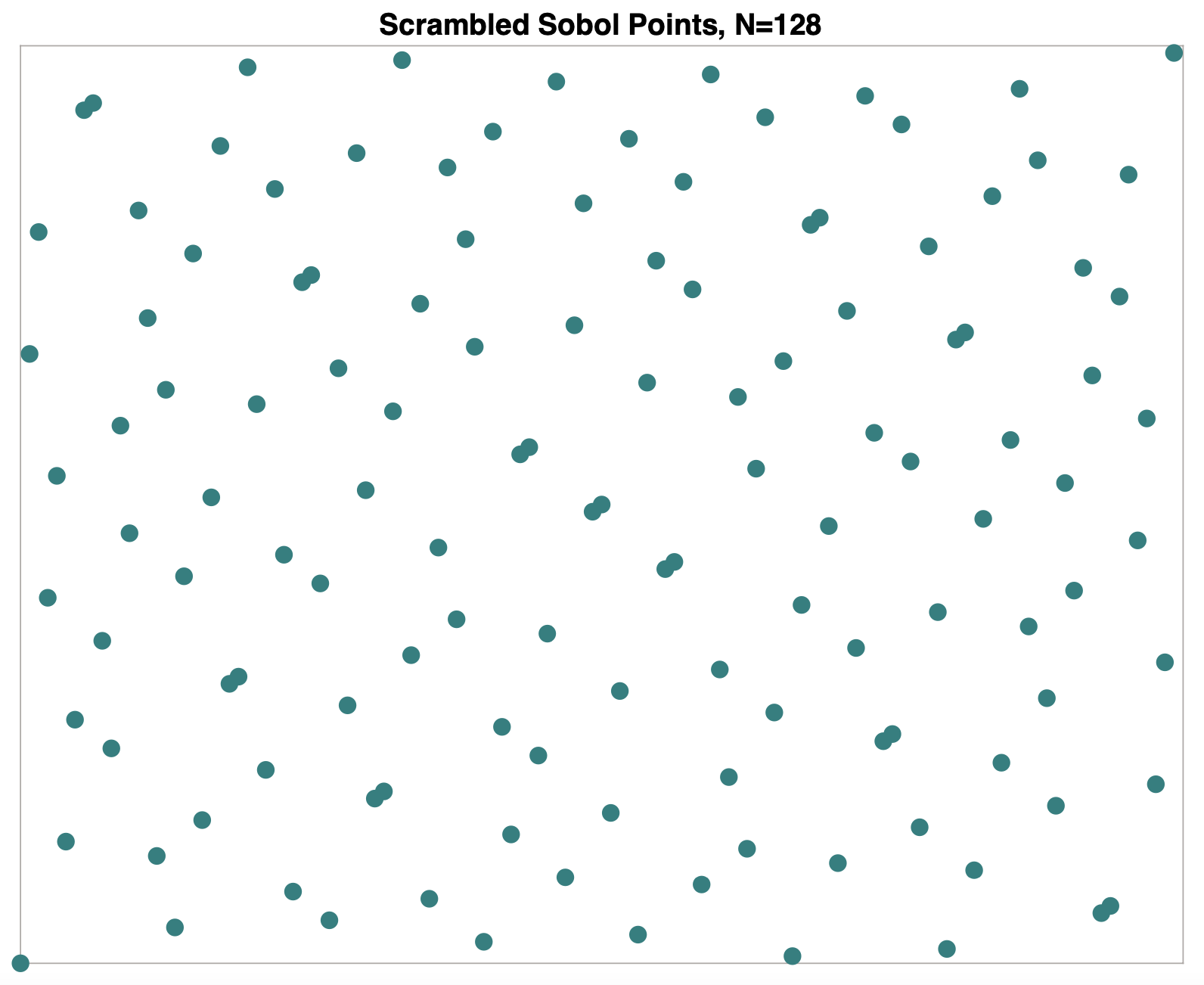}
    \end{minipage}
    \caption{From left to right: MC Points, and QMC Points including Halton, Sobol, and Scrambled Sobol. In each case, the number of points is $N = 128$.}
    \label{fig:mc qmc points}
\end{figure*}

\vspace{-5mm}
\section{Quasi-Monte Carlo Method}

The Monte Carlo method is one of the most widely used numerical methods for simulating probability distributions. It is a decisive step in overcoming the curse of dimensionality problem as computing operations do not grow exponentially but remains manageable and proportional to the dataset's size. A major drawback, however, is that Monte Carlo methods produce probabilistic predictions rather than definitive answers. Another critical point in applying Monte Carlo methods is the generation of random samples. Although Monte Carlo commonly assumes that true random numbers are used as inputs, in practice, they are, in fact, deterministic. These inputs are called pseudo-random numbers and are designed to mimic true random numbers. Quasi-Monte Carlo methods are a deterministic counterpart to the Monte Carlo methods, which
rely on using low-discrepancy sequences with more evenly space-filling properties rather than random sequences \cite{Niederreiter1992}. As a result, there is a significant improvement when using quasi-Monte Carlo for high-dimensional integration to achieve a much faster explicit convergence rate. Furthermore, when a similar problem needs to be solved repeatedly, quasi-Monte Carlo is more reliable, as is often the case in pricing applications, where quasi-Monte Carlo methods have gained tremendous attention \cite{dick2013high,xiang2022quasi,zhang2021efficient}.

Both MC and QMC take the same format when approximating an integral, i.e., 
% \begin{equation}
$\int_{[0,1]^s} f(\bsx) \rd \bsx \approx \frac{1}{N} \sum_{i=1}^{N} f(\bsx_i),$ %\end{equation}
where $\rm \bsx_i \sim U([0,1]^s)$ are independent random samples for MC methods.  In contrast, QMC methods utilize a deterministic low-discrepancy sequence in place of these random samples. The comparison between the two methods is summarized in the table above for quick reference.

\begin{table}
% \caption{Comparison of MC and QMC Methods}
\centering
	\begin{tabular}{ccc}
		\toprule\\[-3mm]
	Monte Carlo  & & Quasi-Monte Carlo \\ 
	\midrule
		%\cline{1-1} \cline{3-3}
		{random}  &&  {deterministic} \\
		$N^{-1/2}$ convergence  &&  close to $N^{-1}$ convergence  \\
		Probabilistic error bounds  && deterministic error bounds \\[1mm]
		\bottomrule
	\end{tabular}
\end{table} \label{tab:mcqmc}

QMC methods  can integrate continuous functions in a hypercube of $d$-dimensional Euclidean space using equal weights with convergence rate $\mathcal{O}\bigl(N^{-1}(\log N)^{d-1}\bigr)$, that is, for a continuous function $f$ on $[0,1]^d$,
\begin{equation}
    \left|\frac{1}{N}\sum_{i=1}^N f(x_i) - \int_{[0,1]^d}f(x)dx\right|\leq C_f\frac{(\log N)^{d-1}}{N},
\end{equation} where ${x_i}_{i=1}^N$ are points from a low-discrepancy sequence and $C_f$ is a constant that depends only on the function $f$.
Roughly speaking, low-discrepancy sequences are designed in such a way that there are no large empty regions or clustered points. The sequence systematically fills gaps in any initial segment, unlike random points. This leads to improved performance compared to traditional Monte Carlo methods.

Recall that codeword $x$ is sent over the AWGN channel, then the conditional target distribution $y|x \sim N(x,\sigma^{2}I_{n})$, where $y_{i}\sim \pi(Y)$. Moving to the QMC way, we will use the low discrepancy point in the sampling procedure for the Gaussian distribution.
Low-discrepancy sequences are a type of sequence widely used in numerical computation and statistical analysis. They include classical constructions such as Halton-type sequences, Sobol points, digital nets\cite{DP2010}, and lattice rules\cite{SloanJoe1994}  along with their variations. The seminal work on low-discrepancy sequences can be found in \cite{Kuipers1974}. When randomization techniques such as random shift or scrambling are applied to these sequences, they become randomized Quasi-Monte Carlo (RQMC) sequences. RQMC estimators are unbiased and can lead to a reduction in variance compared to ordinary Monte Carlo  methods. Furthermore, RQMC provides a probabilistic framework for analyzing QMC-related methods, as discussed in \cite{Lemieux2009} and references therein. The difference between MC and some QMC points is shown in Figure~\ref{fig:mc qmc points}. It is evident that the QMC points are more uniformly distributed within the unit square, as compared to MC points.

Low-discrepancy sequences have also been integrated into Markov Chain Monte Carlo (MCMC) methods, as described in \cite{ChenDickOwen2011, DRZ2013}. The deterministic versions of statistical samplers, such as acceptance-rejection samplers\cite{ZD2016} and importance sampling\cite{DRZ2019}, have been studied, and it has been observed that the use of QMC points can lead to improved performance in various applications.

 \subsection {Using QMC in Importance Sampling}
 In the importance sampling framework, there are at least two ways to improve the algorithm's performance: by designing a more effective proposal distribution and by obtaining better samples from the proposal. This section focuses on the latter aspect, specifically, obtaining better samples from the chosen proposal distribution. To achieve this, the quasi-Monte Carlo method will be used. As a result, we will obtain low-discrepancy samples that are specifically designed for the chosen proposal distribution and have better space-filling properties compared to those obtained using the Monte Carlo method.

The use of QMC points in the importance sampling framework has been explored in previous studies, such as \cite{DRZ2019}, where a significant improvement in numerical integration tasks was observed. More recent research, such as \cite[Theorem 3.1]{He2022}, has shown that using randomized QMC in the importance sampling framework for a Gaussian proposal can lead to an RMSE of the order $N^{-1+\epsilon}$ for arbitrarily small $\epsilon>0$. In the importance sampling framework, QMC points can be used to sample from the proposal distribution, replacing Monte Carlo  points. The higher integration accuracy of QMC methods compared to MC methods means that fewer sampling points are needed to estimate the bit error. However, different types of QMC points may result in slightly different performances. In our experiment, we will consider several commonly used QMC point sets, and further research could be done to construct the optimal QMC/RQMC points for the importance sampling algorithm. This is left for future work.

% \begin{algorithm}
%   \caption{Get low discrepancy sequence for proposal distribution IS}
% %   \begin{algorithmic}

%     \begin{itemize}
%      \item 
%      Generate a low discrepancy sequence U in $[0,1]$ or  $\boldsybol{U}$ in $[0,1]^2$.
%      \item Apply inversion transformation of Gaussian distribution  or box-Muller transformation.
%      \item Carry out importance sampling as in Algorithm 1.
%     \end{itemize}
 
% %   \end{algorithmic}
% \end{algorithm}

% \yg{YG: explain how the proposal is generated by QMC method, and its theoretical estimate similar to MC}

% \section{Adaptive importance sampling by population monte carlo }
% In this section, we introduce the Population Monte Carlo (PMC) framework, which allows us to update the hyper-parameters during simulation. The sampling scheme is unbiased at any iteration and can thus be stopped at any time, thus leading
% to an adaptive importance sampling.

\section{Experiments}
In this section, we demonstrate how different sampling methods perform estimating the bit error in a 6G wireless communication scenario when the number of sampling is large.
We consider only the AWGN channel with the proposed methods. We first randomly generate some code words $x$ and modulate them to latent space. Then we sample $y$ from $g_{\eta}(y \mid x)$  with parameter $\eta$.This is done by adding Gaussian noise based on the proposal distribution on modulated code words. We repeat this process several times. Each time we updated the parameter $\eta$ based on equation \eqref{eq:asymp scaling var}. After we finish sampling, we reweight all the important weights of the samples. Next, we demodulate the noisy signal back and compute the bit error rate for each sampler. The importance sampling estimator is obtained by calculating the weighted sum of the bit error rate. See Algorithm 1.
We evaluate the BER by repeating the simulation multiple times. We compare three methods in the experiment: vanilla Monte Carlo, Quasi-Monte Carlo and importance sampling based on scaling variance. All experiments are simulated in Matlab.

\subsection{MC and QMC points}
In the experiments, we use both Monte Carlo and Quasi-Monte Carlo sampling points. Figure~\ref{fig:mc qmc points} illustrates the MC points by pseudo-random algorithm and QMC points by Sobol, Halton, and scrambled methods from left to right. All the methods sample $n=128$ points in unit square $[0,1]^2$. The MC points in the leftmost are scattered randomly in $[0,1]^2$. 
The QMC points are more regularly aligned in the unit square than MC points but with higher integration accuracy with the same number of sampling points. Sobol and Halton are typical QMC points, while scrambled Sobol points are randomized QMC methods.
Here the ``randomized'' means that each scrambled point is a Sobol point with a small perturbation. The scrambled points have the integration accuracy between MC and QMC methods as mentioned.  

\begin{figure}[t]
    \centering
    \includegraphics[width =0.4 \textwidth]{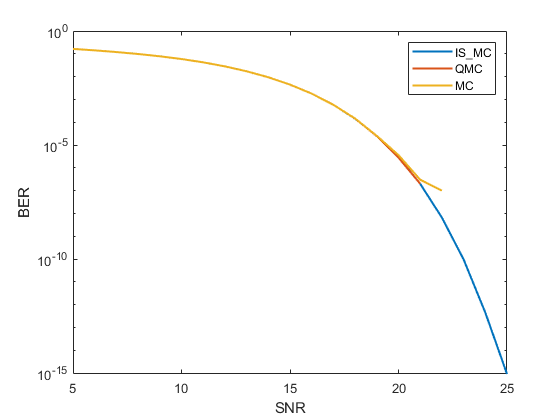}
    \vspace{-3mm}
    \caption{SNR vs. BER for three sampling methods: Monte Carlo, Quasi-Monte Carlo and MC-based Importance Sampling.}
    \label{IS_sampling}
\end{figure}

\subsection{Importance Sampling by Monte Carlo Method}
We present the simulation results of biasing the noise pdf using variance scaling, as well as replacing the Monte Carlo method with the Quasi-Monte Carlo method.

We evaluate the BER for sampling methods for signal-to-noise ratio (SNR, db) from 5 to 25, where the bit error rate is calculated by
% \begin{equation*}
    $\mbox{BER~}  = \frac{\sum_{i=1}^{N} \abs{m_i - m'_i }}{N}$. 
% \end{equation*}
We assume that the input signal $x$ has 5000 codewords and 100 packets, noise $\epsilon$ is Gaussian distributed with mean zero and variance $0.5\times10^{-\frac{\text{SNR}}{10}}$.
Three methods are evaluated in the experiments: the plain Monte Carlo (MC) Method, Quasi-Monte Carlo (QMC) method, and Importance Sampling with MC (IS-MC). In IS-MC, the proposal SNR is the original SNR reduced by 5 using Algorithm~1.

We show the SNR vs. BER in the semilog plot in Figure~\ref{IS_sampling}, where the x-axis is SNR from 0 up to 25, and the y-axis is the BER. 
In the simulation, we use $N_s = 100$ packs and $N = 5000$ packets and qammod order is $16$. The number of sampling points is then $500,000$.
We observe that the importance sampling method achieves great improvement on fixed sampling numbers. When SNR is less than $20$, those three methods show similar performance. However, neither the Monte Carlo method nor the Quasi-Monte Carlo method can estimate BER when SNR is larger than 20. The importance sampling method can precisely estimate BER roughly equal to $10^{-15}$ by using $500,000$ samplings.

\subsection{Comparison of MC and QMC Methods}
We evaluate MC and QMC methods by comparing each of them with importance sampling. As the IS method shows a small variance for each SNR up to 25, we treat the IS result as the ground truth and compute the relative error for the plain MC and the QMC and RQMC methods. The QMC (or RQMC) method provides (nearly) equal-area points which can be used to sample Gaussian distributed points by dividing the norm of each point.
Figure~\ref{Difference of MC, QMC and RQMC} shows the difference of the bit error rates between each of MC sampling, Sobol-based QMC sampling, and scrambled-Sobol-based RQMC sampling with IS-MC sampling.
Both MC and QMC methods have a tiny error when SNR is less than $15$. When SNR is large and the probability of error is small, the variance of both of the two methods increases.
When the number of sampling points is limited (and then SNR is high), the QMC shows a smaller variance than the plain MC method. The RQMC method has a similar performance as the QMC.
\begin{figure}[t]
% \begin{minipage}{0.49\columnwidth}
%     \centering
%     \includegraphics[width = \textwidth]{figure/DIffenerce.png}
%     % \subcaption{The comparison between MC and QMC sampling. The y-axis is the difference of bit errors between MC or QMC with IS-MC.}
%     % \label{Difference between MC and QMC method}
% \end{minipage}
% % \end{figure}
% % \begin{figure}
% \begin{minipage}{0.49\columnwidth}
%     \centering
%     \includegraphics[width = \textwidth]{figure/scramble_difference.png}
% \end{minipage}
\centering
\includegraphics[width = 0.4 \textwidth]{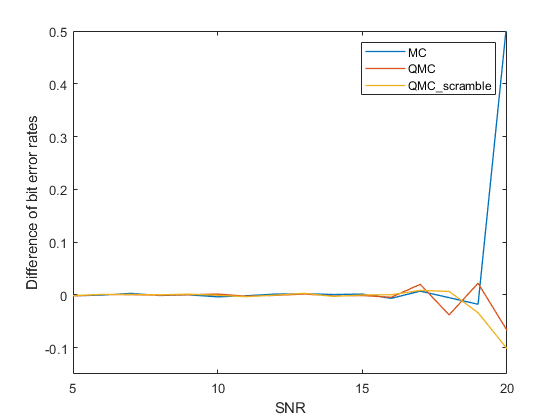}
\vspace{-3mm}
\caption{Comparison of MC with QMC (left), and MC with RQMC (right). The y-axis is the difference of bit error rates between MC, QMC or RQMC with IS-MC. The QMC sampling uses Sobol points and the RQMC is based on scrambled Sobol points.}\label{Difference of MC, QMC and RQMC}
\end{figure}

\section{Conclusion and Discussion}
We propose a new and efficient sampling method for evaluating wireless communication that requires high reliability, specifically for 6G systems. Our approach is based on an adaptive importance sampling algorithm with an updating rule that has a theoretically guaranteed minimal variance for the simplified AWGN case. Our results show that the proposed importance sampling approach is significantly more efficient than traditional Monte Carlo methods, achieving a BER of $10^{-15}$ with 500,000 samplings while Monte Carlo fails at $10^{-6}$. Quasi-Monte Carlo and randomized Quasi-Monte Carlo methods, which have demonstrated higher accuracy in high-dimensional integration, are also potential replacements for Monte Carlo in our importance sampling framework. 

While our main focus is to present a novel concept and framework for simulating rare events in 6G Ultra-Reliable Low-Latency Communications systems, further empirical evaluations comparing the performance and efficiency of our proposed approach to existing methods will be conducted in future work. Moreover, we believe that our framework can be generalized to more complicated scenarios, such as Rayleigh and more general fading channels.

% \section*{Acknowledgment}

% \section*{References}
\bibliographystyle{IEEEtran}
\bibliography{reference}

% \begin{thebibliography}{00}

% \bibitem{b8} 
% Dick, F. Y. Kuo and I. H. Sloan, High dimensional numerical integration - the Quasi-Monte Carlo way. Acta Numerica, 22, 133--288, 2013.
% \bibitem{b9}  J. Xiang and Q. Wang, Quasi-Monte Carlo simulation for American option sensitivities, Journal of
% Computational and Applied Mathematics, 413, 2022.
% \bibitem{b10} C. Zhang, Chaojun, X. Wang and Z. He. Efficient importance sampling in quasi-Monte Carlo methods for computational finance. SIAM Journal on Scientific Computing, 43, 1-29. 2021.

% \end{thebibliography}

\end{document}